\documentclass[twocolumn,showpacs,preprintnumbers,amsmath,amssymb]{revtex4}

\usepackage{graphics}
\usepackage{epsfig}
\usepackage{bm}

\def\jpb{J.\ Phys.\ B: At.\ Mol.\ Opt.\ Phys.\ }

\def\beq{\begin{equation}}
\def\eeq{\end{equation}}

\def\reff#1{(\ref{#1})}

\def\subsc#1{{\mbox{\rm\scriptsize #1}}}

\def\Wcmcm{\mbox{\rm Wcm$^{-2}$}}

\def\Vion{V_\subsc{ion}}

\def\Vxc{V_\subsc{xc}}
\def\Vlaser{V_\subsc{laser}}
\def\Vxlda{V_\subsc{XLDA}}
\def\Nion{N_\subsc{ion}}
\def\Zav{Z_\subsc{av}}
\def\N3d{N_\subsc{3D}}
\def\NKS{N_\subsc{KS}}
\def\HKS{H_\subsc{KS}}

\def\Ninner{N_\subsc{inner}}
\def\xinner{x_\subsc{inner}}
\def\xcm{x_\subsc{c.m.}}
\def\xinnerdach{\hat{x}_\subsc{inner}}
\def\rw{r_\subsc{WS}}

\def\Fint{F_\subsc{int}}
\def\Flaser{F_\subsc{laser}}

\def\vekt#1{\bm{#1}}

\def\vektr{\vekt{r}}
\def\vektx{\vekt{x}}
\def\vektR{\vekt{R}}

\def\vekte{\vekt{e}}
\def\vektE{\vekt{E}}

\def\vektA{\vekt{A}}

\def\vektnabla{\vekt{\nabla}}

\def\halb{\frac{1}{2}}

\def\Edach{\hat{E}}
\def\xdach{\hat{x}}

\def\energy{{\cal{E}}}

\def\pabl#1#2{\frac{\partial #1}{\partial #2}}

\def\imagi{\mbox{\rm i}}
\def\diff{\,\mbox{\rm d}}

\begin{document}
%\draft

\title{Dynamical ionization ignition of clusters in intense and short laser pulses}
\date{\today}
\author{D.~Bauer}
\email[E-mail: ]{bauer@mbi-berlin.de}
\affiliation{Max-Born-Institut, Max-Born-Strasse 2a, 12489 Berlin, Germany}
\author{A.~Macchi}
\affiliation{Dipartimento di Fisica ``Enrico Fermi'', Universit\`a di Pisa \& INFM, sezione A, Via Buonarroti 2, 56127 Pisa, Italy}
\date{\today}

\begin{abstract}
The electron dynamics of rare gas clusters in laser fields is investigated quantum mechanically by means of time-dependent density functional theory. The mechanism of early inner and outer ionization is revealed.  
The formation of an electron wave packet inside the cluster shortly after the first removal of a small amount of electron density is observed. By collisions with the cluster boundary the wave packet oscillation is driven into resonance with the laser field, hence leading to higher absorption of laser energy. Inner ionization is increased because the electric field of the bouncing electron wave packet adds up constructively to the laser field. The fastest electrons in the wave packet escape from the cluster as a whole so that outer ionization is increased as well. 
\end{abstract}


\maketitle

\section{Introduction}
Clusters bridge the gap between bulk material and single atoms. When placed into a laser field all atoms inside (not too big) clusters experience the same laser field, contrary to the bulk where a rapidly created surface plasma and the skin effect prevents the laser from penetrating deeper into the target. In rarefied gases, on the other hand, the laser can propagate but the density is too low to yield high absorption of laser energy and high abundances of fast particles. Hence, the absorption of laser energy is expected to be optimal in clusters. In fact, highly energetic electrons \cite{shao}, ions \cite{ditmirenature,ditmirePRA,springate}, photons \cite{mcpherson,ditmireJPB,teravet}, and neutrons originating from nuclear fusion \cite{zweibackDD} were observed in laser cluster interaction experiments.   
A prerequisite for clusters as an efficient source of energetic particles and photons is the generation of high charge states inside the cluster. Several mechanisms for the increased ionization (as compared to single atoms in the same laser field) have been proposed. Boyer {\em et al.}\ \cite{boyer} suggested that collective electron motion (CEM) inside the cluster is responsible for inner shell vacancies the corresponding radiation of which was observed in experiments \cite{schroeder}. Rose-Petruck {\em et al.}\ \cite{rosepetru} introduced the so-called ``ionization ignition'' (IONIG) where the combined field of the laser and the ions inside the cluster leads to increased {\em inner ionization}, i.e., electrons are more easily removed from their parent ion as if there was the laser field alone. The removal of electrons from the cluster as a whole is called {\em outer ionization}. An ``outer-ionized'' cluster will expand because of the Coulomb repulsion of the ions while a heated, quasi-neutral cluster expands owing to the electron pressure. According to experiments the latter viewpoint of a {\em nanoplasma} (NP) \cite{ditmirePRA,milchberg} seems to be appropriate for big clusters of about $10^4$ atoms or more \cite{kim} while numerical simulations indicate that the Coulomb explosion dynamics prevail for smaller clusters \cite{ishika,rusek}. 
For recent reviews on the  interaction of strong laser light with rare gas clusters see \cite{posthumus,krainsmirn}.

In this paper we are aiming at clarifying the early ionization dynamics in laser cluster interaction. Because the plasma has to be created first the NP model is not applicable. Although we shall find CEM to be indeed an important ingredient for the early ionization dynamics of clusters, the origin of the CEM should be explained rather than assumed beforehand. The IONIG model predicts that ``as the ionization proceeds beyond the first ionization stage, strong electric fields build up within the cluster that further enhance ionization'' \cite{rosepetru}. Here we are interested in how these strong inner fields are generated. Once the cluster is charged after the first electrons are removed, the remaining bound electrons experience the attraction of neighboring ions. This attraction will be largest for bound electrons of ions sitting near the cluster boundary. The other ions will pull these electrons into the cluster interior. This force might be supported by the electric field of the laser, thus leading to further ionization. The IONIG model was successfully called on for interpreting the experimental results in \cite{snyder}.

Although this IONIG mechanism is appealing, the details of the ionization {\em dynamics} remain unclear.
If the force exerted by the other ions is strong enough to ionize further it should be also strong enough to keep the already freed electrons inside the cluster. Hence, it remains to be explained how the electrons are removed from the cluster as a whole (outer ionization). The electrons, after inner ionization, may as well shield the space charge of the ions so that the IONIG mechanism would come to an end, and the NP model would take over even in small clusters. So, why does a strong electric field build up inside the cluster whose interplay with the electric field of the laser is constructive for inner as well as outer ionization?

Classical particle methods are frequently applied to investigate the electron and ion dynamics of clusters in laser fields \cite{rosepetru,lastI,lastII,ishika,saalmann,siedschlagI,siedschlagII} since a full three-dimensional quantum treatment is out of reach with nowadays computers. In these classical accounts inner ionization was either accounted for by sampling the quantum mechanical electron distribution of the bound state by a microcanonical, classical ensemble of electrons, or by using ionization rates so that the electron dynamics was simulated only after inner ionization. Semi-classical Thomas-Fermi theory was employed in \cite{rusek} to study the explosion dynamics of clusters consisting of up to 55 atoms. In our work we use time-dependent density functional theory \cite{rungegross,burkegross,grossdobson}  since we are mainly interested in the early ionization dynamics of clusters were quantum mechanics is important. 

The article is organized as follows. In Section \ref{numerics} the numerical model is introduced. In Section \ref{results} results are presented concerning the groundstate properties of the model cluster (\ref{groundstate}), the electron dynamics (\ref{eldyn}), the formation of collective electron motion inside the cluster and outer ionization (\ref{collouter}), the effect of the collective electron motion on inner ionization (\ref{inner}), the frequency dependence of outer ionization (\ref{frequdepouter}), and the influence of the ionic motion (\ref{ionicmotion}). Finally, we conclude in Section \ref{concl}.

Atomic units (a.u.) are used throughout the paper unless noted otherwise.

\section{Numerical model} \label{numerics}
Time-dependent density functional theory (TDDFT) is employed to study the ionization dynamics of small and medium size rare gas clusters in intense and short laser pulses. To that end the spin degenerate time-dependent Kohn-Sham equation (TDKSE) 
is solved. However, solving the TDKSE in three spatial dimensions (3D) for rare gas clusters in laser fields is too demanding for todays computers. One reason for this is that, contrary to metal clusters \cite{calvay} or fullerenes \cite{bauerc60}, the electrons in the ground state of rare gas clusters are {\em not} delocalized so that jellium models where the ionic background is smeared out and assumed spherical are not applicable.  
In order to make a numerical TDKS treatment feasible two simplifications were made. First, as in previous studies of clusters in laser fields \cite{brewI,brewII,veniard,grigor}, the dimensionality of the problem was restricted to 1D, namely to the direction of the linearly polarized laser field described by the vector potential $\vektA(t)=A(t)\vekte_x$ in dipole approximation \cite{dipoleremark}. 
To that end the Coulomb interactions were replaced by soft-core Coulomb interactions, i.e., $\vert\vektr-\vektr'\vert^{-1} \to [(x-x')^2 + a_{ee}^2]^{-1/2}$ and $\vert\vektr-\vektR_k\vert^{-1} \to [(x-X_k)^2 + a_{ei}^2]^{-1/2}$ for the electron-electron interaction and the electron-ion interaction, respectively. The smoothing parameters $a_{ee}$ and $a_{ei}$ may be chosen to yield ionization energies similar to real 3D systems. The second simplification was the use  of the exchange-only local density approximation (XLDA) so that $\Vxc[n(x,t)]=\Vxlda[n(x,t)]=-\alpha (3 n(x,t)/\pi)^{1/3}$ where $n(x,t)$ is the electron density. The pre-factor $\alpha$ would be unity in full 3D XLDA calculations. In the 1D model it may be chosen to yield satisfactory  ground state properties (we used $\alpha=3/4$).
The ionic potential was $\Vion(x)= -\sum_{k=1}^{\Nion} Z[(x-X_k)^2 + a_{ei}^2]^{-1/2}$ with constant nearest-neighbor distances $X_{k+1}-X_k=d\approx 2\rw$ where $\rw$ is the Wigner-Seitz radius. 
One may look at the 1D ion chain as representing those $\Nion$ ions of a 3D spherical cluster which are situated along the diameter parallel to the linearly polarized laser field. The cluster radius then is $R\approx(\Nion -1) d/2$, and  the number of ions in the real 3D cluster $\N3d=R^3/\rw^3$ would be  $\approx (\Nion-1)^3$. 

The TDKSE for the $\NKS$ Kohn-Sham (KS) orbitals $\Psi_j(x,t)$, $j=1,\ldots,\NKS$ reads 
\beq \imagi\pabl{}{t}\Psi_j(x,t)=\HKS \Psi_j(x,t) \label{tdks}\eeq
where
\beq \HKS  =  -\halb \vektnabla^2 + U  + \Vxlda  + \Vion + \Vlaser.  \label{hks}\eeq
From the doubly spin-degenerate KS orbitals the total electron density $n(x,t)=2\sum_{j=1}^{\NKS} \vert \Psi_j(x,t)\vert^2$ is calculated. 
$U(x,t)=\int\!\!\diff x'\, n(x',t) [(x-x')^2 + a_{ee}^2]^{-1/2}$ is the 1D Hartree potential (accounting for the electron-electron repulsion), and $\Vlaser=-\imagi A(t) \partial_x$ governs the interaction with the laser (taken in velocity gauge with the purely time-dependent term $\sim A^2$ transformed away).
Eq.~\reff{tdks} was solved using the  Crank-Nicolson method with the accuracy of the spatial derivatives boosted to fourth order. Since the Hamiltonian \reff{hks} itself depends (through the density) on the KS orbitals a predictor-corrector method should be used in combination with the usual Crank-Nicolson time-propagation. In practice, however, the evaluation of the Hamiltonian \reff{hks} using the density from the previous time step is usually accurate enough.

For the ion-ion interaction a soft-core potential $Z^2 [(X_j-X_k)^2 + a_{ii}^2]^{-1/2}$ was assumed as well. The ions $j=1,\ldots,\Nion$ were moved according to their classical, non-relativistic equations of motion,
\begin{eqnarray} M\ddot{X_j} &=& Z^2 \sum_{k=1}^{\Nion} \frac{X_j-X_k}{[(X_j-X_k)^2 + a_{ii}^2]^{3/2}} \\ \nonumber 
&& - Z \int\!\!\!\diff x \frac{n(x,t)(X_j-x)}{[(X_j-x)^2 + a_{ei}^2]^{3/2}} \label{ion} \\
&& -Z\pabl{}{t} A(t), \end{eqnarray}
using the Verlet algorithm.

The results which will be presented in the Sections \ref{groundstate}--\ref{frequdepouter} were obtained for an ion mass $M= 131\cdot 1836$ (Xe atom) and $a_{ii}=1$.  Due to the short laser pulse durations ($< 40$\,fs) and the modest charge states $<4$ created, the ionic motion did not appreciably affect the ionization dynamics of our 1D model.  
However, since in a real 3D cluster consisting of $(\Nion-1)^3$ ions the Coulomb explosion would be more violent than in our 1D model with $\Nion$ ions (of the same ion mass and charge state) the issue of ionic motion will be discussed separately in Section \ref{ionicmotion}. 

Finally, before presenting our results, we briefly relate our approach to the work in Refs.~\cite{brewI,veniard,grigor}.
Instead of the simple XLDA, V\'eniard {\em et al.} \cite{veniard} used in their 1D TDDFT approach the more advanced exchange potential proposed by Krieger, Li, and Iafrate (KLI) \cite{kli} but restricted their studies to ion chains up to 7 fixed atoms with one active shell only. Besides the extensive study of the harmonic radiation emitted by the 1D cluster, V\'eniard {\em et al.}  focussed on the dependence of ionization on the internuclear distance and on the number of ions in the chain. This is different from our goal to highlight the dynamics of IONIG. The methodological approach instead is very similar. The only differences are the mobile ions and the choice of simple XLDA instead of the KLI exchange potential. The latter simplification gave us the freedom to simulate bigger clusters and more active shells per atom.

The 1D models in \cite{grigor} and \cite{brewI} were called on to study the explosion dynamics of Xe clusters. Since for this purpose a full TDDFT treatment with many KS orbitals is too demanding even in 1D, further simplifications had to be adopted. In \cite{grigor} a single orbital, representing all electrons, was introduced while in \cite{brewI} time-dependent Thomas-Fermi theory was used.  Our main emphasis in the present paper is complementary to this work since we are not dealing so much with the Coulomb explosion of the cluster but are predominantly interested in the early electron dynamics.

\section{Results and discussion} \label{results}
\subsection{Groundstate properties} \label{groundstate}
Let us consider a chain of $\Nion=9$ ions with nearest-neighbor distance $d=8$  and nuclear charge per ion $Z=4$. Hence, the ground state of the neutral cluster consists of $N=36$ electrons.
The smoothing parameters for the soft-core Coulomb interaction were simply chosen to be  $a_{ee}=a_{ei}=a_{ii}=1$.  
In Fig.~\ref{potsanddens} the ground state density and the various contributions to the total potential are plotted: the ion potential $\Vion$, the Hartree potential $U$, and the exchange potential $\Vxlda$. Although the ionic potential alone has its absolute minimum at the central ion the total potential, including the classical Hartree-repulsion and XLDA, consists of nine almost identical and equidistantly separated potential wells, each locally similar to that of the corresponding individual atom. Consequently, the ground state density displays nine almost equal, well localized peaks, and the energy levels of the cluster are approximately at the same positions as those of the atom, namely around $-1.17$ for the $2\Nion=18$ inner electrons, and around $-0.23$ for the $18$ outer electrons (note, that in the 1D model there are only 2 electrons per shell). Since in XLDA Koopman's theorem is usually not well fulfilled the energy of the highest occupied orbital does not equal the ionization energy for removing the outermost electron. Calculating the ionization energy from the difference of the total energy of the neutral and the singly ionized cluster $\Delta\energy_\subsc{cluster}=0.26$ was obtained. Doing the same for the single atom led to $\Delta \energy_\subsc{atom}=0.49$ which is a reasonable value for rare gas atoms.

\begin{figure}
\caption{\label{potsanddens}  Ground state electron density (orange, dashed-dotted) and the various contributions to the total potential (red, solid): ionic potential $\Vion$ (black, solid), Hartree potential $U$ (blue, dotted), and XLDA potential $\Vxlda$ (green, dashed).}
\end{figure}

\subsection{Electron dynamics}\label{eldyn}
Let us start by comparing the electron motion in the cluster with that of a single atom. In Fig.~\ref{contourcomp} contour plots of the logarithmic density are shown vs.\ space and time. The results for the full cluster (a), the single atom (b), and an artificial cluster made of noninteracting atoms (c) in a laser pulse of frequency $\omega=0.057$ ($\lambda=800$\,nm) and rather modest field amplitude $\Edach=0.033$, corresponding to $\approx 3.9\times 10^{13}$\,\Wcmcm, are shown. The laser field was ramped up linearly over $3$ cycles, held $8$ cycles constant, and ramped down again over $3$ cycles (hereafter called a $(3,8,3)$-pulse) so that the pulse duration was $\approx 37\,$fs. The density in the contour plot (c) was calculated by assuming that all of the $\Nion=9$ atoms behave as the single atom in plot (b).

\begin{figure}
\caption{\label{contourcomp} Logarithmic density $\log n(x,t)$ vs.\ space and time for (a) the cluster with $\Nion=9$, $Z=4$ (b) the single atom, and (c) a cluster made of noninteracting atoms, all with an electron dynamics as shown in (b). The laser parameters were $\omega=0.057$, $\Edach=0.033$, $(3,8,3)$-pulse.    }
\end{figure}

It is seen that the electron dynamics of the cluster (a) and the noninteracting atoms (c) differ significantly from each other already at $t\approx 150$ because more electron density leaves to the right at that time instant in (a) than it does in (c).
This behavior of stronger ionization of the cluster than of the individual atoms continues during the subsequent half laser cycles. 

A qualitatively different electron dynamics inside the cluster emerges for $t\geq 400$. While in the cluster (a) an accumulation of electron density bounces from one boundary of the cluster to the other such an electron dynamics, of course, cannot build up in the ensemble of independent atoms (c). The formation of an electron  wave packet which travels through the entire cluster in step with the laser field is remarkable in view of the fact that the excursion of a free electron in the laser field $\Edach=0.033$ amounts to $\xdach=10.2$ only while the diameter of the cluster is $2R\approx 64$. By varying the cluster size (results for $\Nion=17$ will be presented in the following subsection) it was found that the formation of the bouncing wave packet at laser intensities where $\Edach/\omega^2 < R/2$ is a robust phenomenon, not sensitive to the cluster parameters. Test runs with more advanced exchange potentials such as the Slater potential with self-interaction corrected XLDA and the KLI potential \cite{kli} were performed to ensure that the coherent electron motion is not an artifact of plain XLDA. However, the wave packet formation is sensitive to the laser frequency, as will be shown in subsection \ref{frequdepouter}.

\subsection{Formation of collective electron motion inside the cluster and outer ionization}\label{collouter}
It is useful to study the phase relation between the oscillating electron density inside the cluster and the laser field in order to understand the formation of the electron wave packet with unexpected large excursion amplitude. Given a laser field $\vektE(t)$ of optical (or lower) frequency, the polarization of an atom is $\sim - \vektE(t)$ because the bound electrons are able to follow adiabatically the force exerted by the field. Hence, the phase lag of the polarization with respect to the laser field is $\Delta\phi=\pi$. Free electrons, on the other hand, oscillate perfectly in phase $\sim \vektE(t)$ so that $\Delta\phi=0$. Energy absorption from the laser field is low in both cases because $\int\diff t\, \dot{\vektx}(t) \cdot\vektE(t)\approx 0$ (with $\vektx(t)$ the expectation value for the position of an electron). During ionization there is necessarily a transition from $\Delta\phi=\pi$ to $0$ where energy absorption can take place. During the ionization of atoms this transition occurs rapidly while in clusters, after inner ionization, the electrons may be still bound with respect to the cluster as a whole. Hence, the free motion of electrons $\sim \vektE(t)$ comes to an end at latest when they arrive at the cluster boundary. There, they either escape from the cluster, contributing to outer ionization, or they are reflected so that their phase relation with the driving laser is affected, leading on average to an  enhanced absorption of laser energy. Although collisions of the electrons with ions are included in our TDDFT treatment the effect of the boundary on the electron dynamics clearly dominates. The unimportance of electron ion collisions in medium size and small clusters was pointed out in \cite{ishika} while the relevance of boundary effects was recently affirmed in \cite{megi} within the NP model \cite{remarkII}. 

In Fig.~\ref{contourcomp}a it is seen that some electrons enter about ten atomic units into the vacuum before they are pulled back by the cluster charge. This is reminiscent of what in laser plasma physics is called Brunel effect \cite{brunel}, ``vacuum heating'' \cite{vacheat}, or, more expressively, ``interface phase mixing'' \cite{mulser}.  Thanks to the fact that the fast electrons leave the cluster (and slow electrons are accelerated) a filter effect comes into play so that a wave packet can form that oscillates with the laser frequency and an excursion amplitude of about the cluster radius $R$. 

In order to underpin this scenario the dipole of the electron density inside the cluster $\xinner(t)=\int_{-(R+d/2)}^{R+d/2}\!\!\diff x\, x n(x,t)$ was calculated for several laser and cluster parameters. The results are shown in  Fig.~\ref{phaseplots}. In panel (a) the laser and cluster parameters were the same as in Fig.~\ref{contourcomp}. One sees that during the first few laser periods the electrons indeed move $\sim - E(t)$ (green, dotted curve) as indicated by the first, green bar at the top of the $\xinner$-plot. Then, the electrons inside the cluster get out of phase with the laser for about nine cycles (red bar) so that $\Delta\phi\approx\pi/2$. During this period the dipole amplitude $\xinnerdach$ is particularly high \cite{remark}, and the wave packet bouncing inside the cluster is clearly visible in Fig.~\ref{contourcomp}. Finally, towards the end of the laser pulse the phase relation of the few electrons which were removed from their parent ions but did not make it to leave the cluster becomes that of free electrons (blue bar), i.e., $\xinner(t)\sim E(t)$. In the lower plot the number of electrons inside the cluster $\Ninner$ is plotted and compared with $\Nion$ times the result for the single atom. It is seen that during the first phase (green bar) ionization of the cluster proceeds similar to the single atom case. However, when the phase lag is shifted to $\Delta\phi=\pi/2$ the cluster continues to ionize while the single atom ionization comes to an end.

In  Fig.~\ref{phaseplots}b the same is shown for a higher laser intensity. Essentially, the phase $\Delta\phi$ behaves in the same way but this time, due to the stronger laser field, ionization of the outer shell is almost completed during the first few laser cycles. Hence, the final average charge state is almost the same for the cluster and the single atom. The period where the phase lag is about $\Delta\phi\approx\pi/2$ lasts only a few laser cycles (red bar) and so does the wave packet motion inside the cluster. In Fig.~\ref{phaseplots}c the result for a bigger cluster ($\Nion=17$, $Z=2$) is presented, revealing a qualitatively similar scenario as in (a) with the bouncing wave packet surviving for about 9 cycles. Finally, in panel (d) a higher laser frequency was used ($\omega=0.18$) while keeping the cluster parameters as in (a) and (b). The $\Ninner$-plot reveals that the single atom ionizes more efficiently than the cluster for these laser parameters.
We will come back to the frequency dependence of outer ionization in subsection \ref{frequdepouter}.

\begin{figure}
\vspace{-5mm}
\caption{\label{phaseplots} Dipole $\xinner$ and number of electrons inside the cluster $\Ninner$ vs.\ time for different laser and cluster parameters. (a) $\Edach=0.033$, $\omega=0.057$, $\Nion=9$, $Z=4$ (as in Fig.~\ref{contourcomp}); (b)   $\Edach=0.114$, $\omega=0.057$, $\Nion=9$, $Z=4$; (c) $\Edach=0.114$, $\omega=0.057$, $\Nion=17$, $Z=2$; (d) $\Edach=0.099$, $\omega=0.18$, $\Nion=9$, $Z=4$. The course of $-E(t)$ (a $(3,8,3)$-pulse) is included in the $\xinner$-plots (dotted in green) for distinguishing motion in phase with $-E(t)$ (green bar at the top edge of the panel) and motion approximately $\pi/2$ and $\pi$ out of phase (red and blue bar, respectively). For comparison, $\Nion$ times the result for the single atoms are included in the $\Ninner$-plots. }
\end{figure}

\subsection{Effect of the collective electron motion on inner ionization: dynamical ionization ignition}\label{inner}
The formation of collective electron dynamics as exposed in the previous subsection explains how the absorption of laser energy is increased due to a phase shift into resonance with the driving field, and how the electrons, after inner ionization, are efficiently transported out of the cluster (outer ionization).  The increased inner ionization still remains to be analyzed. IONIG states that the presence of the other ions is responsible for the more efficient removal of bound electrons. This is because two neighboring ions form a potential barrier (cf.\ the $\Vion$ curve in Fig.~\ref{potsanddens}) through which an electron may tunnel when the whole cluster is submitted to an electric field so that the entire potential is tilted. However, we found from our numerical studies that this energetic advantage of a bound electron inside the cluster as compared with an electron in the corresponding single ion is not very pronounced for medium size and small clusters at moderate charge states. Instead, we propose a dynamical version of IONIG where the previously introduced collective electron motion plays an important role. Coherent electron motion was suggested to be responsible for inner shell vacancies in Xe clusters, leading to x-ray emission \cite{mcpherson,schroeder}. In our numerical model there are only two shells and we did not find particularly high line emission from the cluster as compared to the single atom. However, it is possible that, owing to the lack of dynamical correlation in XLDA,  the interaction of the electrons in the wave packet with the still bound electrons is underestimated in our model. Thus, dynamical IONIG might be the mechanism behind the experimental results reported in \cite{mcpherson}.

Despite the fact that in our mean field approach there are no ``hard'' collisions of the electrons in the wave packet with the still bound electrons, the electric field associated with the oscillating electron packet already enhances inner ionization.
In Fig.~\ref{snapshots} two snapshots of the total effective potential and the electron density are presented for the same laser parameters as in Fig.~\ref{contourcomp}. 
In panel (a), electron density and total potential are shown for a time where the electron wave packet is close to the left cluster boundary (the red bar at the bottom of the density plot indicates $\xinner(t)$). 
The contour plot (b) shows for each ion at position $X_i$ the ``difference density'' $\int_{X_i-d/2}^{X_i+d/2} \!\!\!\diff x\ n(x,t)-\Ninner/\Nion$ which indicates whether there is a lack of electron density at that position inside the cluster (black and dark colors) or whether there is excess density (yellow and light colors) compared to the average density $\Ninner/\Nion$.

\begin{figure}
\caption{\label{snapshots} Snapshots of the electron density and the total potential (plus the laser electric field potential alone) at time (a) $t=662.5$ and (c) $t=697.5$. The bar at the bottom of the density plot indicates $\xinner(t)$. Arrows, $+$, and $-$ in the potential plots illustrate the forces $F_\subsc{laser}$, $F_\subsc{int}$  exerted by the laser field and the space charge, respectively. The contour plots (b) and (c) show whether excess (light colors) or lack (dark colors) of electron density (with respect to the average density $\Ninner/\Nion$) prevails at the position of an ion.
The black horizontal lines indicate the times where the snapshots were taken. }
\end{figure}

In panel (a), at time $t=662.5$ the wave packet is close to the left boundary of the cluster while there is a lack of electron density near the right boundary. This charge distribution leads to a force $\Fint$ on the other electrons (pointing to the right) and therefore increases the ionization probability. The electric field of the laser instead is close to zero so that $\Flaser$ is small. The situation approximately a quarter of a laser cycle later is shown in panel (c). The wave packet is at the center of the cluster, moving with maximum velocity to the right and repelling bound or slow electrons in front of it.  The force $\Flaser$ is close to its maximum value at that time, pointing into the direction in which the wave packet moves. The ionization probability is, again, greater than with the laser field alone. Thus, during the course of a laser cycle the total force $\Flaser+\Fint$ clearly leads to higher ionization as if there was the laser field only.

The fact that the electron wave packet dynamics indeed increases inner ionization is underpinned by Fig.~\ref{increasedionization} where the number of electrons in the region $\pm d/2$ around the central ion in the cluster was subtracted from the initial value $Z=4$ and is compared with the corresponding single atom result. The laser parameters were the same as in Figs.~\ref{contourcomp} and \ref{snapshots}. While for the single atom the ionization is completed for $t>700$ the average electron density around the central ion in the cluster is still decreasing. When the electron wave packet sweeps over the central ion the density is temporarily increased, leading to local minima in the curve of Fig.~\ref{increasedionization}. When the wave packet is closest to one of the two cluster boundaries the lack of electrons around the central ion is maximal. The absolute increase of this maximum each half cycle means ongoing inner ionization. In contrast, the single atom, where only the laser field is present but no wave packet can form, does not ionize any further.

\begin{figure}
\quad\hspace{-8mm}%\includegraphics[scale=0.44]{fig5.eps}
\caption{\label{increasedionization} $Z=4$ minus the electron density, integrated $\pm d/2$ around the central ion in the cluster to determine the number of removed electrons as a function of time (solid curve, red).  The result for the single atom is also shown (dotted, black).   }
\end{figure}

\subsection{Dependence of outer ionization on the laser frequency}\label{frequdepouter}
The interaction of the model cluster with $\Nion=9$ was investigated for the two different laser frequencies $\omega_l=0.057$ and $\omega_h=0.18$ (corresponding to $800$ and $254$\,nm, respectively)  and laser intensities between $4\times 10^{12}$ and $10^{16}$\,\Wcmcm. The pulses were of $(3,8,3)$-shape for both frequencies, that is, the pulse durations were $T_l\approx 37$\,fs and $T_h=12$\,fs, respectively.
After the laser pulse, the average charge state in the cluster  $\Zav = Z - \Ninner/\Nion$ was calculated. Note that $\Zav$ only yields information about outer ionization for it does not distinguish between electrons that are still bound to their parent ions and those which move inside the cluster.

\begin{figure}
\quad\hspace{-10mm}%\includegraphics[scale=0.4]{fig6.eps}
\caption{\label{frequcomp}  Average charge state vs.\ laser intensity of the cluster (solid lines) and the individual atom (dotted) for the two different frequencies $\omega_l=0.057$ (lf, drawn red) and $\omega_h=0.18$ (hf, drawn blue).  See text for discussion.   }
\end{figure}

In Fig.~\ref{frequcomp} the average charge state is plotted vs.\ the laser intensity for the two frequencies $\omega_h$ and $\omega_l$. The results for the single atom are also shown. As discussed in the previous subsections, it is seen that in the low frequency case the atoms in the cluster are stronger ionized than an individual atom in the same laser field. Both charge states come close only for $\Zav=2$, that is when the two electrons of the first shell are removed but the two electrons of the next shell are still strongly bound. The stepwise increase of the charge state due to the electronic shell structure is very pronounced in the low-frequency cluster case as well as for the single atoms at both, low and high frequency. The cluster in the high-frequency field instead shows a very different behavior: between $10^{14}\,\Wcmcm$ and the threshold to the inner shell at $\approx 5\times 10^{15}\,\Wcmcm$, the charge state of the single atom is {\em higher} than the average charge state in the cluster.

In Fig.~\ref{eldynhighfrequ} the dynamics of the cluster electrons is shown for a $(3,8,3)$-pulse of frequency $\omega_h=0.18$ and peak field amplitude $\Edach=0.099$, corresponding to an intensity of $3.44\times 10^{14}$\,\Wcmcm. It is seen from Fig.~\ref{frequcomp} that for this intensity the single atom ionizes more efficiently than the cluster as a whole, contrary to what happens at the lower frequency $\omega_l=0.057$. Fig.~\ref{eldynhighfrequ} reveals that the electrons, although removed from their parent ions, mostly remain inside the cluster.  From plot (c) one infers that if the atoms inside the cluster were independent there would be a strong electron emission for $ 100< t < 350$. The emitted electrons have sufficient high kinetic energy to escape from their parent ion (and, thus, from the ``independent atom''-cluster). Contour plot (a), instead, shows that in the real cluster a significant fraction of the electrons near the cluster boundaries return due to the space charge created by {\em all} the ions. A wave packet dynamics that could enhance outer ionization, as in the low frequency result of Fig.~\ref{contourcomp}, does not form. We attribute this to the fact that the initial inner ionization occurs less adiabatic  (multiphoton instead of tunneling ionization). Moreover,   the excursion $\xdach=3.06<d/2$ is too small to trigger any collective motion. The dynamics inside the cluster is rather ``splash-like,'' as can be inferred from the strongly fluctuating electron density between the ions in Fig.~\ref{eldynhighfrequ}a. Hence, although at high laser frequencies inner ionization is high, outer ionization remains low since there is not the coherent electron dynamics supporting outer ionization. Consequently, for creating quasi-neutral nanoplasmas and suppressing Coulomb explosion the use of high frequency lasers is favorable. Reduced ionization of clusters in laser fields of, however,  many times higher frequency (to be generated by x-ray free electron lasers in the near future) was also found in the numerical simulations of \cite{saalmann}. On the other hand, in the soft-x-ray FEL experiment performed by Wabnitz {\em et al.} \cite{wabnitz} {\em increased} ionization of Xe clusters (as compared to single atoms) was observed at $98$\,nm wavelength, $100$\,fs pulse duration, and intensities up to $7\times 10^{13}$\,\Wcmcm. The reason for this unexpected behavior at short wavelengths is not yet clear.

For the high frequency $\omega_h$ and laser intensities $I>3\times 10^{15}\,\Wcmcm$ the average charge state in the cluster overtakes the charge state of the single atom (see Fig.~\ref{frequcomp}). It was therefore interesting to check whether under those conditions a wave packet also forms at the higher frequency $\omega_h$. This is indeed the case. However, due to the rapid ionization the wave packet dynamics lasts only a few laser cycles (or less).

\begin{figure}
\caption{\label{eldynhighfrequ} Same as in Fig.~\ref{contourcomp} but for the higher frequency $\omega_h=0.18$ and $\Edach=0.099$. Logarithmic density $\log n(x,t)$ vs.\ space and time for (a) the cluster, (b) the single atom, and (c) a cluster made of noninteracting atoms.   }
\end{figure}

\subsection{Influence of ionic motion on the wave packet formation} \label{ionicmotion}
So far the ion mass was set to $M=131\cdot 1836$ (Xe atom). This high mass lead to no appreciable ionic motion in the 1D cluster during the pulse. Even in the ``worst case'' where all active electrons are removed at $t=0$ so that the ionic charge $Z=4$ is not screened at all, the $\Nion=9$-cluster radius increases only from $32$\,a.u.\ to $33.8$\,a.u.\ within the 37\,fs pulse duration. However, the corresponding spherical 3D cluster with $(\Nion-1)^3$ Xe-ions of charge state $Z=4$ doubles its initial radius within the same time \cite{doublingformula}. Clearly, the Coulomb explosion is underestimated in the 1D model because in 3D an ion sitting at the cluster surface ``sees'' not only the charges of the ions aligned along the laser field polarization direction but essentially a charged sphere containing all the other $(\Nion-1)^3-1$ ions. 

Hence, in order to allow the 1D ion chain to explode similarly to the corresponding 3D cluster one has to reduce the ion mass. To obtain the results shown in Fig.~\ref{mobileions} $M=1836$ was set. In the ``worst case'' introduced in the previous paragraph the radius of the 1D cluster now expands from $32$ to $\approx 130$\,a.u.\ within $37$\,fs. However, in the early stage when the wave packet formation inside the cluster takes place the charge states of the ions are still low and the cluster radius remains close to its original value. This is clearly visible in Fig.~\ref{mobileions}a where between $t=300$ and $t=800$ the bouncing wave packet can be easily identified. In Fig.~\ref{mobileions}b,c  the dipole $\xinner$ and the number of electrons inside the cluster $\Ninner$ are presented, as in Fig.~\ref{phaseplots}. For comparison, the results for a computer run with immobile ions is included (blue and broken curves). The differences are small. Only the ionization at the end of the laser pulse is slightly reduced in the case of mobile ions. This is expected because the ions gain their kinetic energy at the expense of the electrons. 

At higher laser intensities the higher charge states are generated earlier. In these cases the Coulomb explosion is more violent but the wave packet mechanism contributes little to the total ionization of the cluster anyway. In Fig.~\ref{mobileions_strongfield} the ionization and Coulomb explosion scenario for the highest laser intensity $I=1.14\times 10^{16}\,\Wcmcm$ considered in this paper is presented for the same cluster parameters as in Fig.~\ref{mobileions}. The outer shell of all atoms is depleted rapidly already during the first laser cycle both for the cluster and the isolated atom. On average three quarters of the electrons in the next shell are depopulated during the remainder of the laser pulse in the case of the cluster with mobile ions. The ionization probability of the isolated atom is slightly less. The ionization of the cluster with heavier ions is, again, higher. 

Due to the rapidly increasing charge states of the ions no formation of an electron  wave packet moving $\Delta\phi=\pi/2$ out of phase with respect to the driving laser field for several laser cycles can be inferred in Fig.~\ref{mobileions_strongfield}a. 
Instead, the typical motion $\sim \vektE(t)$ of free electron density in the laser field $\vektE(t)$ seems to dominate the dynamics. It is interesting to observe that the removal of the electrons in the first shell is even more efficient for the isolated atom than in the cluster (see Fig.~\ref{mobileions_strongfield}c around $t=180$). However, shortly after ``cracking'' the second shell the cluster ionization overtakes the single atom result. At that time  the $\xinner$-plot in  Fig.~\ref{mobileions_strongfield}b reveals two extrema (marked by arrows) which are approximately $\Delta\phi=\pi/2$ out of phase with the driving field. Thus, although a dephased wave packet dynamics over several cycles cannot develop in such intense fields the stronger ionization  of the cluster still relies on the fact that the electron density inside the cluster moves temporarily with the appropriate phase lag necessary for efficient absorption.

\begin{figure}
\caption{\label{mobileions} Result for $\Edach=0.062$, $\omega=0.057$, $\Nion=9$, $Z=4$, (3,8,3)-pulse, and mobile ions ($M=1836$). (a) Logarithmically scaled electron density, (b) $\xinner$, and (c) $\Ninner$ (as in Fig.~\ref{phaseplots}). The blue and broken curves in (b) and (c) are the results for immobile ions.  }
\end{figure}

\begin{figure}
\caption{\label{mobileions_strongfield} The same as in Fig.~\ref{mobileions} but for $\Edach=0.57$.  }
\end{figure}

\section{Conclusion} \label{concl}
The ionization dynamics of a one-dimensional rare gas cluster model in intense and short laser pulses was investigated by means of time-dependent density functional theory. An electron wave packet dynamics was found to build up inside the cluster when the laser intensity $I=\Edach^2$ was sufficiently high for modest inner ionization but  not that high that all electrons of an atomic shell are freed within a few laser cycles. The electron wave packet is driven into resonance with the laser field through the collisions with the cluster boundary. The phase lag between the bouncing electron wave packet and the laser field then is $\pi/2$ so that the absorption of laser energy is particularly high. The fastest electrons in the wave packet escape from the cluster (outer ionization).  The electric field of the bouncing electron wave packet adds up constructively to the laser field, thus enhancing inner ionization. This effect was called {\em dynamical ionization ignition}. 
It is a robust phenomenon with respect to the cluster size and, since it occurs during the early ionization stage, it is not affected by ionic motion. However, with increasing  laser frequency (keeping the laser intensity fixed) the mechanism is less efficient.

We expect the wave packet scenario being valid also in real, three-dimensional rare gas clusters. Due to the spherical cluster-vacuum boundary the wave packet should assume a sickle-like shape in that case. However, in order to verify this, studies with higher-dimensional cluster models will be pursued in the future.

\begin{acknowledgments}
This work was supported by the Deutsche Forschungsgemeinschaft (D.B.) and through the INFM advanced research project Clusters (A.M.). The permission to run our codes  on the Linux cluster at PC$^2$ in Paderborn, Germany, is gratefully acknowledged.
\end{acknowledgments}

%%%%%%%%%%%%%%%%%%%%%


\begin{thebibliography}{}
\bibitem{dipoleremark} The dipole approximation can be applied because the wavelengths of the laser light under consideration are many times greater than the cluster diameters in our study, and magnetic field effects do not play a role.
\bibitem{remarkII} On the other hand, experimental results could several times successfully be interpreted by employing the NP model where collisional absorption is supposed to be essential \cite{lezius,springate}. Whether the success of the NP model in this context is just by coincidence or collisional absorption is really important is not yet clear.
\bibitem{remark} Note, however, that $\xinner(t)$ is the dipole and does not describe the center of the oscillating wave packet. The center of mass of all inner electrons is $\xcm=\xinner/\Ninner$. However, to $\xinner$, $\xcm$, and $\Ninner$  also the electrons that are still bound to their parent ions contribute so that $\xcm\ll R$ holds despite the fact that the oscillating wave packet has an excursion $\approx R$.
\bibitem{doublingformula} The time during which the cluster radius doubles reads  $t_d=1.6\,[MR^3/(\Nion-1)^3Z^2]^{1/2}$ (cf.\ Eq.\ (42) in \cite{krainsmirn}).
\end{thebibliography}
\end{document}